\documentclass[aps,prl,amsmath,amssymb, preprintnumbers, reprint,longbibliography,superscriptaddress]{revtex4-1}
\pdfoutput=1
\usepackage{lmodern, graphicx, multirow, xcolor, adjustbox}
\usepackage[colorlinks=true, citecolor=blue, urlcolor=blue, linkcolor=blue, breaklinks=true, pdfpagelabels=false]{hyperref}

\makeatletter\g@addto@macro\bfseries{\boldmath}\makeatother
\DeclareMathSymbol{\shortminus}{\mathbin}{AMSa}{"39}

\catcode`\$=\active
\gdef$#1${\texorpdfstring{\(#1\)}{\detokenize{#1}}}

\usepackage{lipsum}
\usepackage{multirow}
\usepackage{verbatim}

\usepackage{cancel}
\usepackage{hyperref}

\begin{document}

\newcommand{\cpthree}{Centre for Cosmology, Particle Physics and Phenomenology (CP3), Universit\'{e} catholique de Louvain, 1348 Louvain-la-Neuve, Belgium}

\author{C\'eline Degrande}
\email{celine.degrande@uclouvain.be}
\affiliation{\cpthree}

\author{Matteo Maltoni}
\email{matteo.maltoni@uclouvain.be}
\affiliation{\cpthree}

\preprint{CP3-20-58}

\title{Reviving the interference: framework and proof-of-principle for the anomalous gluon self-interaction in the SMEFT}

\begin{abstract}
Interferences are not positive-definite and can change sign over the phase space. If the contributions of the regions where they are positive and negative nearly cancel each other, their effects can be hard to measure. In this paper, we propose a method to quantify the ability of an observable to separate these opposite contributions and therefore to revive the interference effects at experiments. We apply this strategy to the anomalous gluon operator in the SMEFT, for which the linear-term suppression is well known. We show that we can get, for the first time, constraints on its coefficient from the interference only that are similar to those from the square of the new-physics amplitude.
\end{abstract}

\maketitle

\subparagraph{Introduction}
The Standard Model Effective Field Theory (SMEFT) explores the deviations in the Standard Model (SM) couplings due to new states that are too heavy to be directly produced at the LHC or any other current experiment. Nonetheless, those new states can affect the interactions between the SM particles, and accurate measurements of their strengths should reveal or constrain the presence of unknown particles.

In this framework, heavy new degrees of freedoom are integrated out and the new physics is parametrised by higher-dimensional operators $O_i$ with coefficients $C_i$ \cite{Grzadkowski:2010es,Buchmuller:1985jz}, that are included in the Lagrangian as
\begin{equation}
\mathcal{L}_{\small{SMEFT}} = \mathcal{L}_{\small{SM}} + \sum_{i} \frac{C_i}{\Lambda^2}\hspace{1mm}O_i + \mathcal{O}(1/\Lambda^4),
\end{equation}
where $\Lambda$ is a cutoff scale. Observables such as differential cross sections display the same expansion
\begin{equation}
\frac{d\sigma}{dX} = \frac{d\sigma^{SM}}{dX} + \sum_i  \frac{C_i}{\Lambda^2}\frac{d\sigma^{1/\Lambda^2}}{dX}+ \mathcal{O}(1/\Lambda^4),
\end{equation}
where $X$ is a generic measurable variable. While constraints should ideally come from the second term, {\it i.e.} the linear one in the coefficients, in practice they are often obtained from the quadratic order in $C_i$. This is mainly due to the fact that the linear term is an interference between the amplitudes with one insertion of a SMEFT operator and the SM one, and this has been shown to be suppressed in $2 \to 2$ processes \cite{Azatov:2016sqh}. As it will be illustrated below, the same occurs in higher-multiplicity processes as well. 

A suppression of the total $\mathcal{O}(1/\Lambda^2)$ cross section can have two origins: either the interference matrix element is small all over the phase space, or it changes sign over different portions of it. This letter aims, in the second case, to restore the linear term using differential measurements and to assess the efficiency of the reviving procedure.

Although we will focus on a single operator in the rest of the letter, the method is generic and can be applied to any cancellation due to a sign flip over the phase space, even outside the SMEFT. Other obvious applications in the SMEFT are the CP-violating operators: their interference with the SM does not contribute to the total cross section of C-even processes, but it can be probed using CP-violating observables \cite{Degrande:2021}.

\begin{table}
\caption{\small{LO $\mathcal{O}(1/\Lambda^2)$ cross sections, in pb, and percentages of positive-weighted events for some processes with a non-null interference between the SM and the $O_G$ operator. $C_G$ is set to 1 and $\Lambda$ to 5 TeV. Different minimum cuts on the light-jets transverse momenta are considered. The total transverse energy divided by two ($H_T/2$) is chosen as dynamical scale. Numerical uncertainties are of order of a few percent}} \label{tab:processes}
\begin{tabular}{c|cc|cc|cc}
\multicolumn{1}{c}{} & \multicolumn{2}{|c}{$p_T^j > 50$ GeV} & \multicolumn{2}{|c}{$p_T^j > 200$ GeV} & \multicolumn{2}{|c}{$p_T^j > 1000$ GeV} \\
 & $\sigma$ & w$>$0 & $\sigma$ & w$>$0 & $\sigma$ & w$>$0 \\
\hline
$t\bar{t}$ & 1.384 & 85\% & 1.384 & 85\% & 1.384 & 85\% \\
$t\bar{t}j$ & 5.1$\cdot 10^{-1}$ & 62\% & 1.17$\cdot 10^{-1}$ & 60\% & 1.44$\cdot 10^{-3}$ & 62\% \\
$jjj$ & 3.7$\cdot 10^1$  & 52\% & 7.3$\cdot 10^{-1}$  & 52\% & 4.55$\cdot 10^{-4}$ & 61\% \\
$jjjj$ & -3.1$\cdot 10^1$ & 45\% & -1.53$\cdot 10^{-1}$ & 44\% & -3.8$\cdot 10^{-6}$& 39\% \\        
\hline
\end{tabular}
\end{table}

\begin{table*}
\caption{\small{LO cross sections, in pb, for three-jet production with different values of the minimum $p_T^j$ cut. $C_G$ is set to 1, $\Lambda$ to 5 TeV and the renomalisation and factorisation scales are fixed to the reported values. Only up to one $O_G$ insertion is allowed. The percentages of positive-weighted events are shown for the interference, together with the $\sigma^{|\text{meas}|}$ and $\sigma^{|\text{int}|}$values}} \label{tab:3j_xsect}
\begin{tabular}{cc|c|cccc|c}
    & & SM & \multicolumn{4}{c|}{$\mathcal{O}(1/\Lambda^2)$} & $\mathcal{O}(1/\Lambda^4)$ \\ 
   min $p_{T}^j$ (GeV) & $\mu_{R,F}$ (GeV) & $\sigma$ & $\sigma$ & wgt$>0$ & $\sigma^{|\text{meas}|}$ & $\sigma^{|\text{int}|}$ & $\sigma$ \\ \hline
   50 & 150 & 9.73$\cdot 10^5$ & 1.5$\cdot 10^1$ & 50.4\% & 7.81$\cdot 10^2$ & 1.051$\cdot 10^3$ & 3.922$\cdot 10^1$    \\
   200 & 500 & 8.96$\cdot 10^2$ & 4.6$\cdot 10^{-1}$& 51.4\% & 8.77  & 1.251$\cdot 10^1$ & 2.737   \\
   500 & 1000 & 3.11 & 1.87$\cdot 10^{-2}$& 54.0\% & 1.508$\cdot 10^{-1}$ & 2.243$\cdot 10^{-1}$ & 1.484$\cdot 10^{-1}$ \\
   1000 & 2000 & 9.08$\cdot 10^{-3}$& 4.58$\cdot 10^{-4}$& 60.1\% & 1.470$\cdot 10^{-3}$ & 2.297$\cdot 10^{-3}$ & 3.062$\cdot 10^{-3}$ \\ \hline
\end{tabular}
\end{table*}

\subparagraph{Framework}
In this work, we focus on the dimension-6 operator 
\begin{equation}
   O_G = g_S^{} f^{abc} \hspace{1mm} G^{a,\mu}_{\nu}G^{b,\nu}_{\rho}G^{c,\rho}_{\mu},
\end{equation}
with $G_{\mu\nu}$ the gluon field strength. While it is expected to contribute to multijet and top-pair production, its interference vanishes for dijet and is strongly suppressed for other processes. Previous studies~\cite{Dixon:1993xd, Krauss:2016ely, Hirschi_2018} indeed suggest that a good sensitivity to its interference is unachievable. However, constraints on this operator are essential as it affects, for example, top-quark production \cite{Buckley:2015lku}.

High-multiplicity jet measurements can set strong limits on this operator, but so far these are mainly derived from the $\mathcal{O}(1/\Lambda^4)$ terms \cite{Krauss:2016ely, Hirschi_2018}. The large number of dimension-8 operators makes the complete evaluation of this order tricky to compute. Throughout this paper, we truncate the amplitude at $\mathcal{O}(1/\Lambda^2)$ level, therefore the $\mathcal{O}(1/\Lambda^4)$ term only includes the square of the $1/\Lambda^2$ amplitudes. The most stringent bound on this operator comes from the $\mathcal{O}(1/\Lambda^4)$ contribution to dijet measurements and reads $C_{G}/\Lambda^2 < 0.031$ TeV${}^{-2}$ at 95\% confidence level (CL) \cite{Goldouzian:2020wdq}.

We use the TopEffTh \cite{topeffth} Universal FeynRules Output (UFO) \cite{Degrande:2011ua,Alloul:2013bka} as a model for our study. All the operator coefficients are set to zero but the $O_{G}$ one, which is taken equal to 1 with $\Lambda=5$ TeV. {\sc MadGraph5}\_a{\sc MC@NLO} v2.8.2 \cite{Alwall_2014} is then employed to generate LO events at parton level for the SM, the square of the $\mathcal{O}(1/\Lambda^2)$ amplitude and their interference. The QED contributions are ignored. All the quarks but the top are considered as massless. For the light jets, a minimum $\Delta R_{jj}$ of 0.4 and maximum pseudorapidity $|\eta|$ of 5 are requested, and different minimum $p_T^j$ cuts are considered. Multiple insertions of the dimension-6 operator are not allowed. We use the NNPDF2.3 parton distribution function (PDF) set \cite{Ball:2012cx} and the results are presented for the LHC at 13 TeV. We leave NLO corrections, parton shower (PS) and detector effects for future studies.

The cancellation of the linear term is efficient if the integrals of the squared amplitude for the interference in the phase-space regions where it is positive or negative are almost equal in absolute value. Those two integrals are obtained from the sum of the weights of Monte Carlo events generated according to the interference, keeping respectively only positive- or negative-weighted events. Throughout this paper, we always assume the generated events to be unweighted.

In Table \ref{tab:processes}, we use the percentage of positive weights to quantify the efficiency of this cancelation for some top and jet processes that are affected by $O_G$. In three-jet production, this value is the closest to 50\% and the cross section is large enough for accurate differential measurements, so in the remaining of this letter we will focus on this process and leave the others for future analyses. In each three-jet sample we fix both the renormalisation and factorisation scales $\mu_{R,F}$ to the same value, depending on the cut imposed on the $p_T$ of light jets: the numbers are shown in Table \ref{tab:3j_xsect}.

We define the ``integrable cross section'' as
\begin{equation}
   \sigma^{|\text{int}|} = \int d\Phi \left| \frac{d\sigma^{1/\Lambda^2}}{d\Phi} \right| = \lim_{N\to\infty} \sum_{i=1}^{N} |w_i|.
\end{equation}
It is computed from the sum of the absolute values of $N$ weights $w_i$ and represents the total interference effect over the whole phase space $\Phi$. This quantity is given in Table~\ref{tab:3j_xsect} together with the SM, linear and quadratic total cross sections in the three-jet case. The comparison between $\sigma^{|\text{int}|}$ and the total linear cross section shows that a large cancellation occurs over $\Phi$ among positive- and negative-weighted contributions: contrarily to dijet production, the interference might be restored in three-jets to obtain competitive bounds on $C_G/\Lambda^2$.

It is important to notice, though, that $\sigma^{|\text{int}|}$ is not a measurable quantity, as it requires the knowledge not only of the momenta of the jets and of the incoming partons, but also of their flavours and helicities. Therefore, we define the ``measurable integrable cross section'' as
\begin{equation}
   \sigma^{|\text{meas}|} = \int d\Phi_\text{meas} \left| \sum_{\{\text{um}\}}\frac{d\sigma^{1/\Lambda^2}}{d\Phi} \right|,
\end{equation}
where $\{\text{um}\}$ is a set of unmeasurable quantities, that depends on the process and the collider. The sum can be sometimes replaced by integrals over continuous unmeasurable quantities, such as the longitudinal momentum component of a neutrino. This variable quantifies the difference between the positive and negative contributions to the interference cross section using only the information that is experimentally available. It can thus be considered as an upper bound for any asymmetry built on one or a few kinematic variables aiming at restoring the interference, and can be used to assess the efficiency of such asymmetry. $\sigma^{|\text{meas}|} $ is estimated, over a sample of $N$ Monte Carlo events, as
\begin{equation}
   \sigma^{|\text{meas}|}  = \lim_{N\to\infty} \sum_{i=1}^N w_i \cdot \text{sign}\left(\sum_{\{\text{um}\}} \text{ME}(\vec{p}_i,\{\text{um}\}) \right),
\end{equation}
where ME is the part of the squared amplitude due to the interference and $w_i$, $\vec{p}_i$ label the weight and the momenta of the jets of the $i^{th}$ event. Therefore, this can be seen as a matrix-element method at the partonic level to revive the interference \cite{Kondo:1988yd,Kondo:1991dw,Kondo:1993in,Dalitz:1991wa,Dalitz:1992np,Goldstein:1992xp,Dalitz:1992bx}. The values of $\sigma^{|\text{meas}|}$ for three-jet production are given in Table \ref{tab:3j_xsect}, for different cuts on $p_T^j$. To compute them, for each event we assign to the incoming (outgoing) momenta all the possible permutations of the initial- (final-) state quark flavours for all the three-jet subprocesses, and check the sign of the sum of the squared amplitudes that we obtain. We assume perfect momenta reconstruction, as no neutrinos are present and PS effects are ignored.
It can be seen that the cancellation among positive and negative weighted events decreases with the $p_T^j$ cut, while the ratio $ \sigma^{|\text{meas}|}/\sigma^{|\text{int}|}$ remains roughly constant.

\begin{figure}
\caption{\small{Differential cross-section distributions for the leading jet $p_T$ ({\it top}) and the transverse sphericity ({\it bottom}) in three-jet production, with $p_T^j > $ 200 GeV. The red (blue) histograms represent the differential cross-section contributions to the interference by the positive- (negative-) weighted events. Their differences, in orange, are the differential linear terms; the dotted portion in the bottom plot is the opposite of the negative part. The black lines reproduce the SM distributions divided by 100, while the green lines show the squared ones. The last bin of the top plot contains the overflow}} \label{fig:distrib}
   \includegraphics[width=0.49\textwidth]{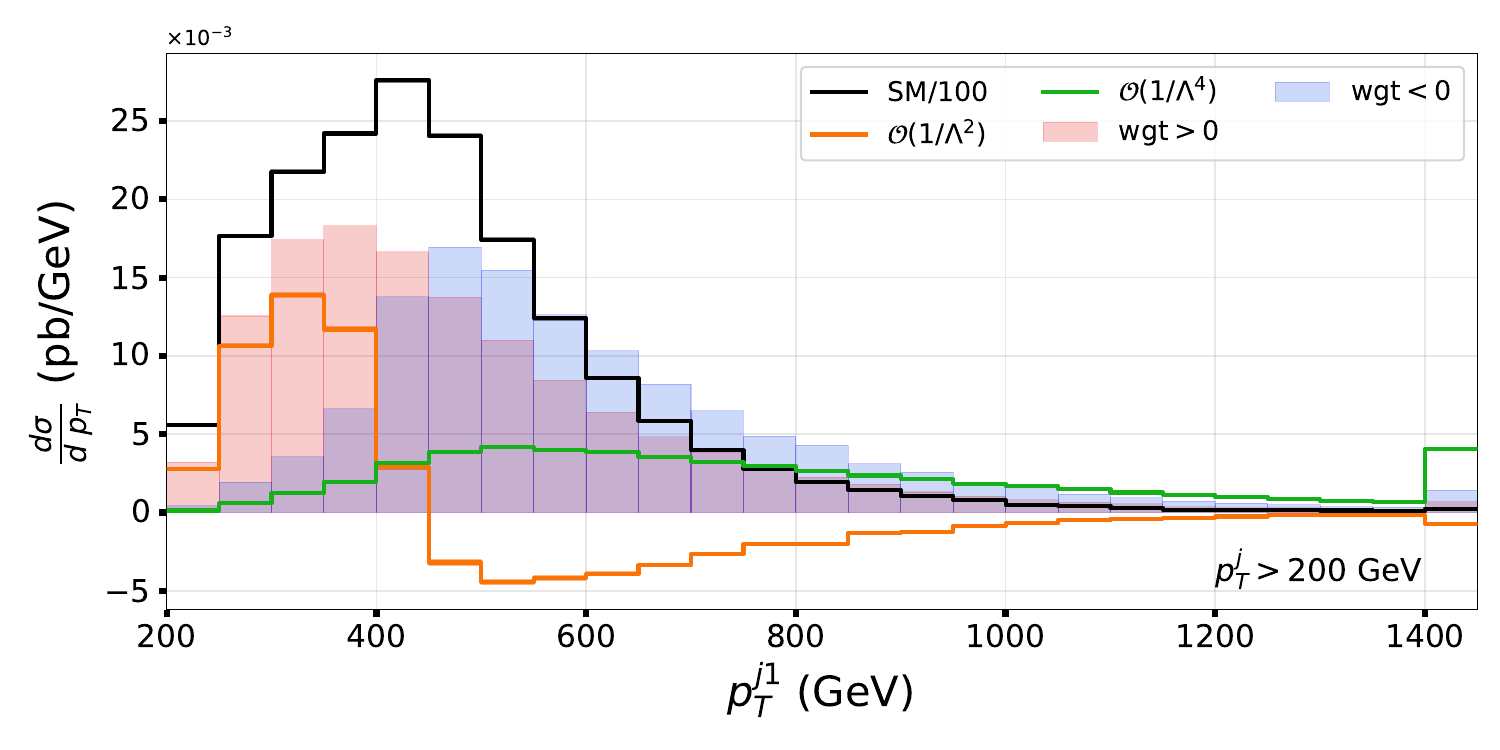} 
   \includegraphics[width=0.48\textwidth]{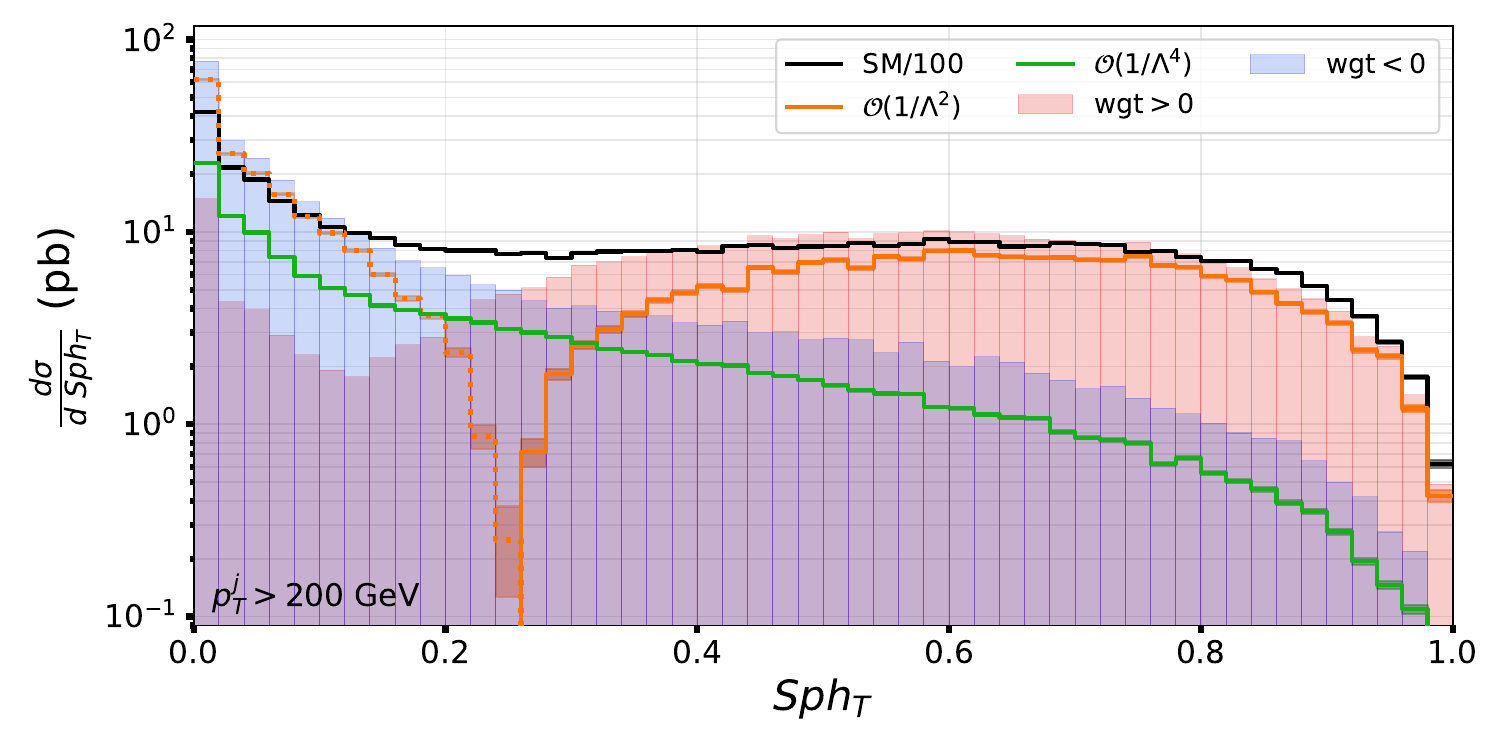} 
\end{figure}

\subparagraph{Differential distributions}
The comparison of $\sigma^{|\text{meas}|}$ and $\sigma^{|\text{int}|}$ confirms that a large part of the actual interference effect can be accessed for this process at the LHC. In principle, the measurable integrable cross section represents the best observable to restore the linear term, but it is computationally costly because of the squared-amplitude calculation.

For this reason, we test the ability to separate positive and negative weights for various differential and double-differential cross sections, comparing their efficiencies against the $\sigma^{|\text{meas}|}$ value. The considered distributions include the transverse momenta $p_T$ and the pseudorapidities $\eta$ of the jets, their angular distances $\Delta R$, their invariant masses, the normalised triple product among the three-momenta of the jets, and some event-shape variables, including the transverse thrust, the jet broadening \cite{Khachatryan:2014ika} and the transverse sphericity \cite{Banfi:2010xy}. 

Several observables, such as the $p_T$ of the leading jet $p_T^{j1}$ and the angular distance between the two lowest-$p_T$ jets $\Delta R_{j2 j3}$, achieve an efficiency up to 50\% with respect to $\sigma^{|\text{meas}|}$. For comparison, the efficiency of the total cross section is about 5\%, as it can be seen in Table \ref{tab:3j_xsect}. The best efficiency, however, is obtained from the transverse sphericity and is above 80\%. The transverse thrust presents a similar value. Moreover, these numbers barely change with the global lower cut on each of the three jets $p_T$. The transverse sphericity $Sph_T$ is defined from the eigenvalues $\lambda_1 \geq \lambda_2$ of the transverse momentum tensor, namely
\begin{equation}
   M_{xy} = \frac{1}{ \sum_i |\vec{p}_{T,i}| } \sum_{i=1}^{N_\text{jets}} \frac{1}{ |\vec{p}_{T,i}| } \left( 
              \begin{array}{cc} 
                 p_{x,i}^2 & p_{x,i}p_{y,i} \\ p_{y,i}p_{x,i} & p_{y,i}^2 
              \end{array} \right),
\end{equation}
through the relation
\begin{equation}
   Sph_T = \frac{2\lambda_2}{\lambda_2 + \lambda_1}.
\end{equation}
This variable is often used to differentiate events that are back-to-back ($Sph_T \sim 0$) from those that feature a more isotropic distribution of the jets momenta ($Sph_T\sim1$). These two topologies contribute with opposite signs to the linear-term cross section, leading to the suppression we observe. This explains why the phase-space cancellation is less effective with higher $p_T$ cuts, as a strong hierarchy between the jets becomes unlikely. The differential distributions for the $p_T$ of the leading jet and $Sph_T$ are shown in Fig.~\ref{fig:distrib}, with the separate positive- and negative-weighted contributions. Contrarily to other less efficient variables, the two opposite-sign plots have different trends, resulting in a non-zero distribution for the interference that changes sign across the phase space.

The transverse thrust shows similar reviving effects as the sphericity. Moreover, it is a linear function of the momenta and, thus, infrared- and collinear-safe. The same cannot be said of $Sph_T$, but the lower cuts on $p_T^j$ avoid any strong dependence on the hadronisation.

NLO predictions for the interference of operators known for their suppression over the phase space seem to lead in general to very large and/or negative $K$-factors, as it is the case for the $O_W$ operator, the electroweak (EW) analogous of $O_G$ \cite{Degrande:2020evl}. This can be understood as the result of a cancellation between regions contributing positively and negatively to the interference, each of them singularly presenting more reasonable but different $K$-factors. If this is the case, only observables that are able to separate the opposite-sign regions could yield stable predictions and NLO bounds for the interference. Due to the heavy computation needed, we leave the NLO corrections to our observables for future work.  

\begin{table*}
\caption{\small{95\% CL bounds on $C_G /\Lambda^2$ in TeV${}^{-2}$ for different cuts on the jets $p_T$, at linear and quadratic level for three-jet production. The number of bins is reported for each distribution. The $Sph_T$ cut is the value, between 0 and 1, at which we separate the two bins used for the transverse sphericity. $S_T$ is defined in \cite{Krauss:2016ely}}} \label{tab:best_bounds}
\begin{tabular}{c|ccc|cc}
   min $p_T^j$ (GeV) & Distribution & $Sph_T$ cut & $N_\text{bins}$ & $\mathcal{O}(1/\Lambda^2)$ bounds & $\mathcal{O}(1/\Lambda^4)$ bounds \\ \hline
   50 & $p_T^{j3}$ vs $Sph_T$ & 0.23 & 12 & [-1.5, 1.5] & [-4.1, 4.7]$\cdot 10^{-1}$ \\
   200 & $S_T$ vs $Sph_T$ & 0.25 & 32 & [-2.9, 2.9]$\cdot 10^{-1}$ & [-1.0, 1.0]$\cdot 10^{-1}$ \\
   500 & $M_{j2j3}$ vs $Sph_T$ & 0.31 & 32 & [-8.4, 8.4]$\cdot 10^{-2}$ & [-4.4, 6.1]$\cdot 10^{-2}$ \\
   1000 & $M_{j2j3}$ vs $Sph_T$ & 0.35 & 22 & [-3.1, 3.1]$\cdot 10^{-2}$ & [-1.7, 2.3]$\cdot 10^{-2}$ \\
   \hline
\end{tabular}
\end{table*}

\begin{figure}
\caption{\small{ 95\% CL upper bounds on $\Lambda$ (for $C_G = 1$) as functions of the upper cut over the CoM energy $\sqrt{s}$, inferred from the best distribution for the 200 and 1000 GeV $p_T^j$ cut. The orange (green) lines show the bounds from the linear (quadratic) order. The dotted lines reproduce the limits obtained through the $S_T$ variable alone. The axes on top quantify the percentage of events, in the interference sample, that are discarded because of the cut on $\sqrt{s}$. The shaded areas cover the region where $\sqrt{s}$ is larger than the bound on $\Lambda$ }} \label{fig:lambda_bounds}
   \includegraphics[width=0.48\textwidth]{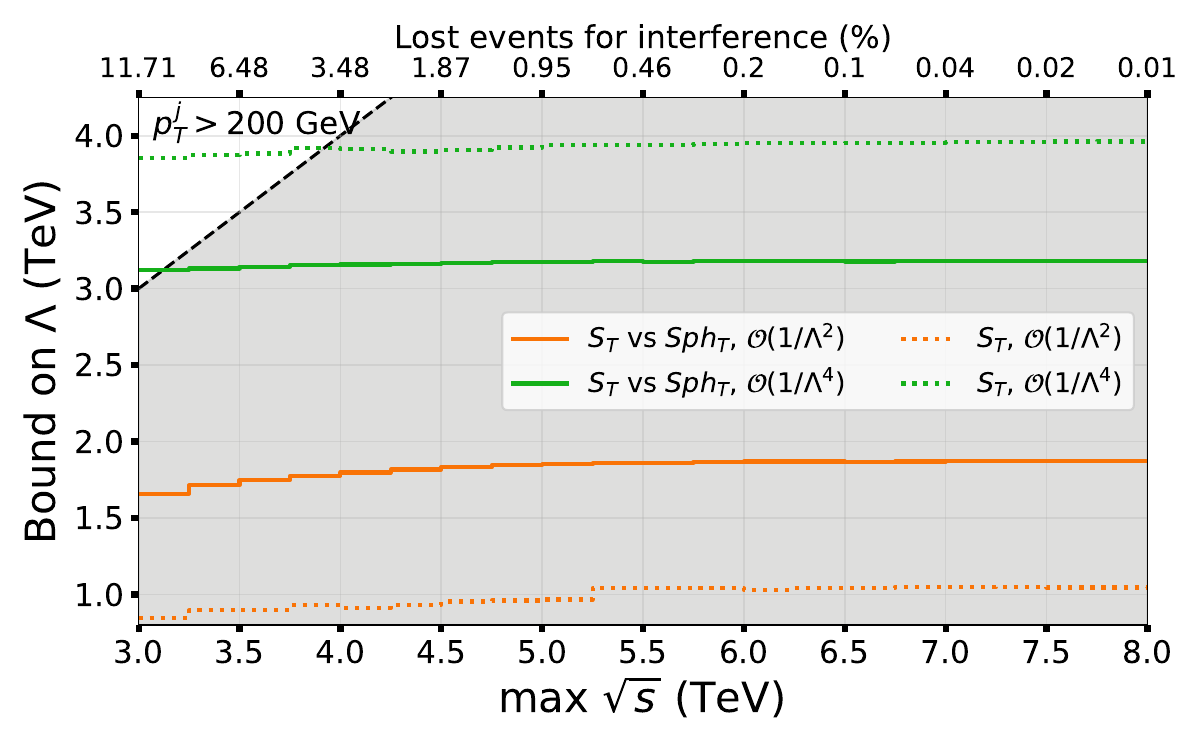} 
   \includegraphics[width=0.48\textwidth]{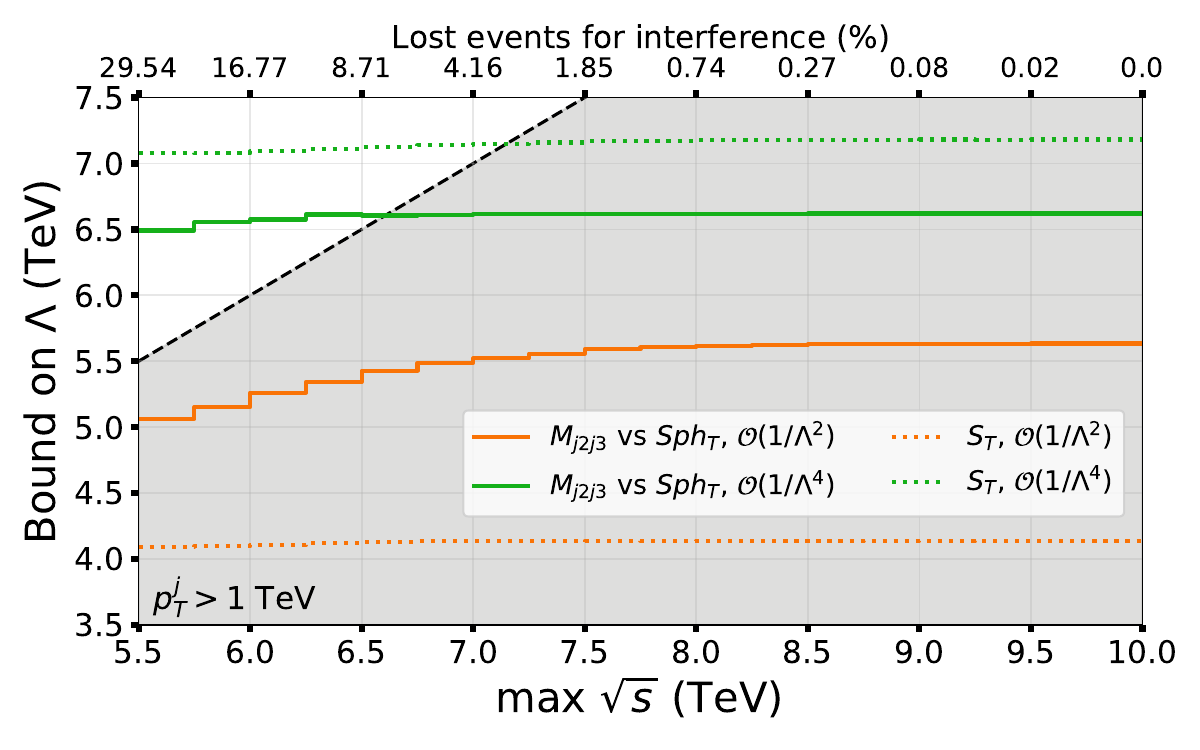} 
\end{figure}

\subparagraph{Limits on $C_G/\Lambda^2$}
Using the transverse sphericity to split the positive and negative contributions, we can estimate the limits that could be obtained on $C_G /\Lambda^2$ from the interference only or including the squared contribution as well. The bounds are computed, for each distribution, through the $\chi$-square expression
\begin{equation}
\chi^2 = \sum_{i=1}^{N_\text{bins}} \left( \frac{ x_{i}^\text{exp} - x_{i}^\text{th} }{ \Delta_i } \right)^2
       = \sum_{i=1}^{N_\text{bins}} \left( \frac{C_G}{\Lambda^2}\hspace{1mm} \frac{ x_{i}^{1/\Lambda^2} }{ \Delta_i } \right)^2, \label{chisq}
\end{equation}
where $x_{i}^\text{exp}$ and $x_{i}^\text{th}= x_{i}^\text{SM}+ \frac{C_G}{\Lambda^2} x_{i}^{1/\Lambda^2} \left( +\frac{C_G^2}{\Lambda^4} x_i^{1/\Lambda^4} \right)$ are respectively the measured and predicted contents of each bin. Since the experimental results for the distributions we are interested in have not been published yet, we assume that the measured data will follow the SM for the considered quantities, resulting in the last step of Eq.~\eqref{chisq}. The uncertainty $\Delta_i$ in each bin is assumed as 10\% of its SM content; this estimate seems consistent with available experimental results \cite{Aad:2014rma}. We choose our binnings so to ensure that the numerical errors in our simulated data are below this threshold.

The best results are displayed in Table \ref{tab:best_bounds}. These most stringent limits come from double-differential distributions of $Sph_T$ with some observables that grow with the energy, depending on the $p_T^j$ cut: the transverse momentum $p_T^{j3}$ of the third-leading jet, the invariant mass $M_{j2 j3}$ of the second and third jets, and the transverse energy $S_T$ of all the jets with $p_T > 50$ GeV, defined in \cite{Krauss:2016ely}. For the transverse sphericity, only two bins are employed between 0 and 1: their border is placed where the sign flip occurs in its differential cross sections, and the values are reported in the table. The results we obtain from the linear term are comparable to the quadratic-level ones and to the bounds already present in the literature, for the highest $p_T^j$ cuts.

Finally, to assess the validity of the SMEFT with our approach, we display in Fig.~\ref{fig:lambda_bounds} how the limits on $\Lambda$ vary if different cuts on the center-of-mass (CoM) energy are applied, assuming $C_G=1$. In principle, the EFT is valid if $\sqrt{s}<\Lambda$, which is not satisfied even at upper cuts of few TeV, for this choice of $C_G$. The constraints barely change when the events with $\sqrt{s} \gtrsim 8$ TeV are included in the $p_T^j > 1$ TeV case. The bounds obtained through the interference grow faster than the ones which involve the $\mathcal{O}(1/\Lambda^4)$ term too, as it is expected from their different dependency on $\Lambda$. The limits that come from the $S_T$ variable alone are also shown, for comparison.

\subparagraph{Conclusions}
We used the sign of the measurable matrix element as a tool to revive the interference and to quantify the efficiency of differential distributions to separate negatively- and positively-contributing regions of the phase space in three-jet production. We found efficient distributions that can lift the interference suppression for anomalous gluon interactions with the SM, and managed to set contraints over the $O_G$ operator coefficient at linear level that are compatible with the ones from the squared term which are already present in the literature. In particular, event-shape observables like the transverse sphericity seem to be particularly sensitive to this suppression in multijet processes. Therefore, we have found an answer to the long-standing quest for a sensitivity to the interference between the anomalous gluon operator and the SM; more studies are still needed to explore other processes and to include NLO corrections and PS effects. As an example, the similar magnitude of the three- and four-jet cross sections in Table \ref{tab:processes} suggests that a proper matching and merging procedure should be employed to get reliable results. Due to its perceptiveness to the interference, $Sph_T$ can also provide information on the sign of $C_G$. Finally, the proposed strategy can be easily reinterpreted in other BSM scenarios outside the SMEFT, as it relies on purely kinematic distributions: it is fully generic and can be applied to any suppression due to sign changes over the phase space.

\subparagraph{Acknowledgements}
We are grateful to Fabio Maltoni and Vincent Lemaitre for interesting discussion during the completion of this analysis. This work was funded by the F.R.S.-FNRS through the MISU convention F.6001.19.

\bibliography{refs.bib}

\end{document}